\documentclass[10pt,a4paper]{article}
\usepackage{amsmath,amsfonts}
\usepackage{algorithmic}
\usepackage{array}
\usepackage[caption=false,font=normalsize,labelfont=sf,textfont=sf]{subfig}
\usepackage{textcomp}
\usepackage{stfloats}
\usepackage{url}
\usepackage{verbatim}
\usepackage{graphicx}
\usepackage{xcolor}
\usepackage{subcaption}
\usepackage{tabularx,booktabs}
\usepackage[sort&compress, numbers]{natbib}
\usepackage[textfont=rm, textfont=footnotesize, labelformat=default,labelfont=footnotesize,labelsep=quad,skip=3pt]{caption}
\def\BibTeX{{\rm B\kern-.05em{\sc i\kern-.025em b}\kern-.08em
    T\kern-.1667em\lower.7ex\hbox{E}\kern-.125emX}}
\usepackage{balance}
\begin{document}
\providecommand{\newblock}{}
\newcommand{\3}{\textsubscript{3}}
\newcommand{\2}{\textsubscript{2~}}
\newcommand{\cc}{\textsubscript{c~}}
\newcommand{\ccc}{\textsubscript{c}}
\newcommand{\rt}{R\textsubscript{T}}
\newcommand{\rl}{R\textsubscript{L}}
\newcommand{\Tmax}{T\textsubscript{max}}
\newcommand{\Est}{E\textsubscript{st}~}
\newcommand{\Vc}{V\textsubscript{coil}~}

\title{Numerical analysis of non-insulated DEMO TF coils}
\author{{M. Ortino, N.Bykovskiy, X.Sarasola, P.Bruzzone, K.Sedlak}
\thanks{This work has been carried out within the framework of the EUROfusion Consortium, via the Euratom Research and Training Programme (Grant Agreement No 101052200 — EUROfusion) and funded by the Swiss State Secretariat for Education, Research and Innovation (SERI). Views and opinions expressed are however those of the author(s) only and do not necessarily reflect those of the European Union, the European Commission, or SERI. Neither the European Union nor the European Commission nor SERI can be held responsible for them. (Corresponding author: Mattia Ortino). 
The authors are with the École Polytechnique Fédérale de Lausanne (EPFL), Swiss Plasma Center (SPC), CH-5232 Villigen PSI, Switzerland (e-mail: mattia.ortino@epfl.ch).
} } 
\maketitle

\begin{abstract}
The design of small solenoids without/with-partial insulation among turns or layers has been proven to grant higher thermal stability than for standard insulated cases. This technology implies higher safety standards, by allowing for 1-100 V highest voltage in the single TF coil; at the same time, a passive/simpler quench protection systems (QPS) becomes a concrete opportunity.
Besides, experiments and simulations on large-scale solenoids do not guarantee yet the same advantages, as the redistribution of currents after quench is not yet clear and the temperature margins allowed by LTS are narrower. The possibility for full internal energy dissipation is here discussed analytically, highlighting the already high final temperatures being reached and the limits of the held hypotheses. \\ 
This study presents therefore a new effective 3-D original numerical model, developed for simulating in a fast fashion charge/discharge and quench of a EUROfusion DEMO toroidal field (TF) coil, by coupling both electrical and thermal physics in the same solver. Its structure is explained as well as its first benchmarks. Finally, the first results are reported of a parametric study where only turn-turn localized bridges are considered, discussing the consequences and the future outlook for this simulation work. 
\end{abstract}


\vspace{-5mm}
\section{Introduction} \label{intro}
New high-energy projects involving superconducting magnets as fusion reactors (e.g. SPARC, BEST, EU-DEMO) or particle accelerators (e.g. FCC-hh, CEPC, Muon collider) call for a revision of the coil design, as these facilities foresee 15/20 T-class magnets.\\
This involves many details to be reconsidered, as for the chosen superconductor (SC) in the winding pack (''WP'', now very often built with HTS \cite{hartwig2020, gao2023, fabbri2024, pong2024}), structural support (casing) \cite{Wang2023}, insulation system and therefore quench protection strategy. Particularly in fusion coils, the need for fast charging/discharging of the pulsed coils (e.g. central solenoid modules \cite{martovetsky2002,oh2002,wu2017}, as well as poloidal field coils \cite{ogasawara1984,heller1996,darweschsad1996}) declared so far insulation systems to be essential for any kind of magnet.  In this respect, organic materials as Kapton or glass fibre with epoxy/wax/cyanate ester have so far been the used standard for SC - mainly LTS - cables. They also prevent against any transversal heat propagation, e.g. during a quench.
Unfortunately, organic insulations also result in the mechanical weak-point of the whole magnet structure, being much softer than the surrounding Cu/SC/stainless steel (SS). This comes from the different elastic behaviour on mechanical strain ("sponge effect") \cite{evans2001}, which cannot be avoided and limits the maximal allowed coil temperatures during a quench (e.g. 150 K for DEMO TF). A failure could be also due to radiation resistance issues of the resins \cite{weber2011, wu2013}, these being further stressed with next generation machines (see DEMO, FCC-hh) foreseeing higher dose levels \cite{reis2023,prioli2019}. Another drawback is the very high voltages developed at the coil terminals V\textsubscript{coil} (e.g. 1-30 kV during quench), which can lead to severe faults (e.g. arcs) during operation and maintenance \cite{hamada2023,arendt1981}. Non-Insulated (NI) and Partial-Insulated (PI) technologies goal is to mitigate if not eliminate these problems, by introducing contact resistances between the coil turns/layers, eventually counting on a complete passive protection.\\
As for self-protection characteristics, LTS magnets are on one side better secured against quench than HTS, as the propagation velocity of the quenched zone (NZPV) is much lower in HTS due to their higher specific heat (1-100 cm/s vs 1-50 m/s Nb\textsubscript{3}Sn or 10-80 m/s NbTi): the steep temperature rise will affect only a small volume of the whole winding, not redistributing the heat-load over significant lengths. On the other side, HTS have much bigger critical surfaces, making the thermal stability margin much higher than for LTS coils.
In this respect, a trending wave of small NI/PI coils were built and successfully tested since 2010 \cite{hahn2010, wang2013, uglietti2014}. The vast majority of these attempts use coated conductors (HTS), as for their high temperature stability but also for the intrinsic geometrical feature of the HTS tapes, where controlled resistance in a layer/pancake winding can be provided by direct contact between the Cu flat surfaces (``metal-insulation'' MI, \cite{lecrevisse2022, hahn2018}) following the simple equation:
\begin{equation}
\label{Rvsrho}
R(T)=\rho(T)~l_{bridge}/A_{bridge}
\end{equation}
\noindent where \textit{l\textsubscript{bridge}} is the length of the contact resistance while \textit{A\textsubscript{bridge}} its surface. These coils show very high self-protection features against quench  and over-current (very low or null burning rates), as a certain amount of current from the power supply I\textsubscript{p.s.} can radially by-pass its original (superconducting) spiral path through turn-to-turn contact. Some excess of over-currents (transient) I\textsubscript{trans }can develop with non-negligible amplitudes so that I\textsubscript{trans}$\gg$I\textsubscript{p.s.}, eventually requiring a transient stress analysis because of the high associated mechanical forces. Tuning the contact resistance towards higher values seems to mitigate the over-current amplitudes, still conceiving a fast quench propagation but allowing for static stress analysis \cite{markiewicz2019}. All these evidences, as well as the first successes on compact-size fusion coils \cite{vieira2024}, let believe this technology to be the path to next-gen high(-er) field magnets, with the declared highest goal of removing the external energy damp strategy with all its complex associated problems.\\
In a tokamak fusion reactor, only the TF WP seems suitable for applying NI/PI technologies, as these coils are envisaged for operating at stable DC conditions and not foreseeing frequent charges/discharges. Besides the coil shape and performance of the selected conductor (Sec. \ref{intro}), also the coil mass plays an important if not crucial role for a proper insulation design.  \\
In this respect., it is possible to analytically calculate the maximum temperature T\textsubscript{max} the coil would reach - e.g. after a quench event - if its stored energy E\textsubscript{stored} is dissipated inside the same coil mass everywhere at the same time (isothermal WP) via Eq.\ref{EvsM}:
\begin{equation}
\label{EvsM}
m_{conductor} \int_{T_0}^{T_{max}} c_p(T) dT=E_{stored}
\end{equation}
\noindent where T\textsubscript{0} is 4.2 K and T\textsubscript{max} the highest temperature reached in the coil after the energy is completely released. \\
Here c\textsubscript{p}(T) is the effective heat capacity of the winding only or could as well account for some of the surrounding metal casing (the higher the SS \% contributing to thermal facts, the lower will be \Tmax). These numbers are displayed for few coil examples in Tab~\ref{Tab1}, where the SPARC TF coil final temperatures \Tmax~are given as a range as their precise material partition in the WP is not precisely known.\\
\begin{table}
\caption{Energy stored \Est and average, final temperatures \Tmax for different types of coil (\cite{hartwig2020, kircher1999,vieira2024, sborchia2008, bottura2001})}
\label{Tab1}
\begin{tabular}{| c | c | c | c | c |}
\hline
Coil  & Stored Energy E\textsubscript{st} & Total conductor mass [kg] & Energy density [J/kg] & ~\Tmax[K] \\ 
\hline
15 T Lab Solenoid & 250 kJ  & 50 & 5*10\textsuperscript{3}& $\simeq$70 \\
\hline
LHC Dipole & 8 MJ & 800 & 10*10\textsuperscript{3} &  $\simeq$100  \\ 
\hline 
SPARC TFMC &  110 MJ & 5.1*10\textsuperscript{3}  &  22*10\textsuperscript{3} & $\simeq$130-170\\
\hline
 SPARC TF  & 316 MJ  & 8*10\textsuperscript{3}  & 40 *10\textsuperscript{3}& $\simeq$230-270\\
  \hline 
 CMS Solenoid  & 2.6 GJ  & 220*10\textsuperscript{3}  &  11*10\textsuperscript{3}& $\simeq$80 K \\
  \hline 
 ITER TF  & 2.28 GJ  & 43 *10\textsuperscript{3} & 40*10\textsuperscript{3} & $\simeq$300\\ 
 \hline
\end{tabular}
\vspace{-7mm}
\end{table}
Tab~\ref{Tab1} shows also the energy densities (J/kg) for the different cases. In general terms:  $\simeq$2000 J/~kg will lead to \Tmax= 50 K,  $\simeq$10.000 J/kg to 100 K,  $\simeq$60.000 J/kg to 300 K. Approaching NI/PI designs for GJ-class \Est coil can be therefore prohibitive, unless the latter is provided with a very big mass (often not helping the design) or very long cooling times are acceptable. 
\section{DEMO TF: circuit analysis} \label{DEMO TF: analytical background}\label{analytical}
\begin{figure}[t]
\centering
\includegraphics[height=0.8\textwidth,width=0.5\textwidth]{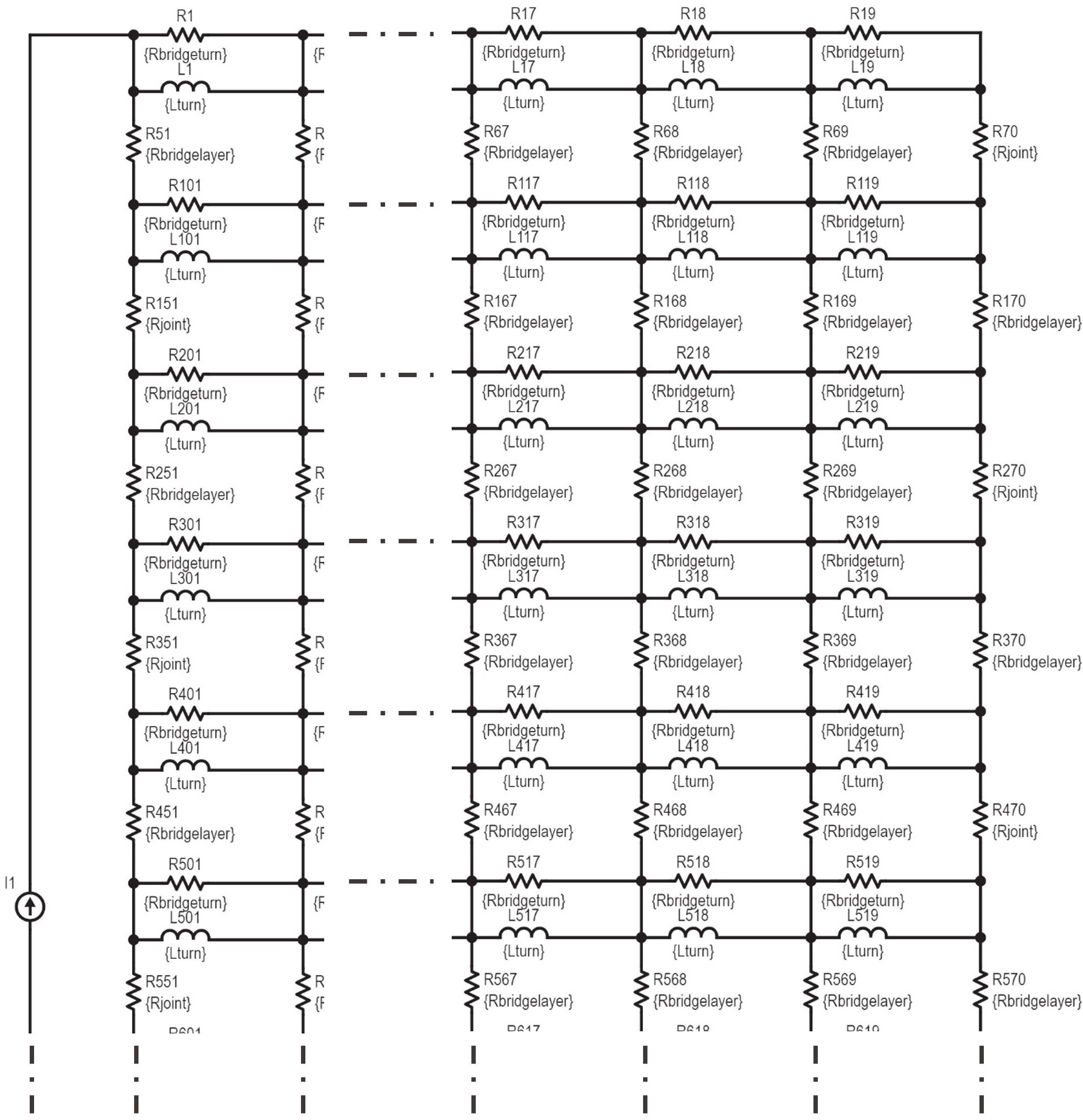}
\caption{Cropping of the DEMO TF (12 T, 66 kA, 12 layers, 19 turns) equivalent electric model used for a parametric study conducted with LTSpice\textregistered.}
\vspace{-3mm}
\label{Spice}
\end{figure}

The hypotheses hold in the last analyses (isothermal WP) are well known to respond to ideal cases, as the coil volume involved in a quench propagation is usually much smaller than the total. Moreover, for NI/PI coil cases, the mutual relation between shared current and Joule heating (produced by the currents shared through the Cu, SS and contact resistances) is much more complex and can lead to unexpected results. As for DEMO TF, purely electrical parametric analyses were performed at SPC with LTSpice\textregistered~\cite{OrtinoMT_2023, SarasolaASC_2022}. The latter aimed at identifying the appropriate turn-turn bridge resistance R\textsubscript{T} and layer-layer bridge resistance R\textsubscript{L} by solving transient analysis on a circuit as depicted (cropped) in Fig. \ref{Spice}, where $R_{bridgelayer}$=R\textsubscript{L} and $R_{bridgeturn}$=R\textsubscript{T}; I\textsubscript{p.s.}=66 kA, H\textsubscript{coil}=3.55 H and $R_{joint}$=1 n$\Omega$. A list of conditions for coil charge/discharge was set as input parameters: leak current I\textsubscript{leak}$\le$0.01 I\textsubscript{p.s.}, longest acceptable charging time t\textsubscript{charge}=30.000s and delay time - between I\textsubscript{p.s.} and whatever I in the coil - to be t\textsubscript{delay}$\le$6000s.
As a result: R\textsubscript{L} $ \ge$ 2 m$\Omega$, and R\textsubscript{T} =5 $\mu\Omega$-10$\mu\Omega$.\\
Since relying on circuit models neglects any thermal fact essential for analysing a quench scenario, an original code in MATLAB was developed in order to couple thermal and electrical physics, with the aim of delivering - in a fast fashion - info on a) T\textsubscript{min}/ \Tmax~reached at the end of a transient (charge or quench) and b) total coil voltage \Vc, as well as resistive V\textsubscript{R} and inductive V\textsubscript{L} developed in each branch of the simulated circuit as a function of temperature.

\section{Electro-thermal MATLAB\textregistered~code} \label{code}
\subsection{Code structure}\label{structure}
\noindent The script is a distributed lumped model, meant to first reproduce a coil geometry by tuning the amount of layers/pancakes and turns following the user requirements, then to simulate either the electrical and/or thermal evolution due to the Joule heating (or additional heat sources).\\Starting from a 1-D configuration where the conductor length is displayed as lines connecting the nodes N, it is possible to specify the materials surface within the cable cross-section, resulting in a simple 2-D rendering. On a node, the conductor cross-section is isothermal, while heat exchange happens only among the nodes, therefore meaning both along and across the conductor path. \\
 Finally, the code allows to establish simple relations between any N - also along the Z axis - resulting overall in a 3-D solver. 
 \begin{figure}[htbp]
\centering
\includegraphics[width=0.8\textwidth]{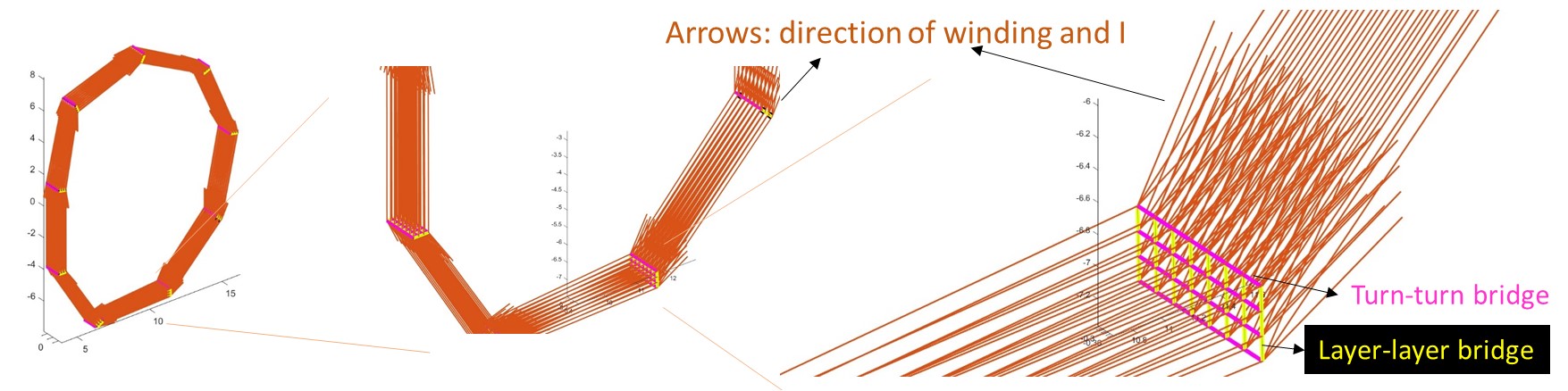}
\caption{Rendering of a DEMO TF-coil with 10 N x 12 layers x 19 turns (2280 total nodes: the orange arrows indicate the path the current takes; on the right, a zoom is provided showing turn-to-turn and layer-to-layer contact bridges across the winding pack;}
\label{Geometry}
\end{figure}
\noindent Its visual interface is displayed as in the example of Fig. \ref{Geometry}, where the direction of the winding and path of the currents are showed with orange lines. In this case, the DEMO TF geometry is built by using N =10  (found to be the minimum acceptable for representing the shape of the winding) x 12 layers x 19 turns (as from the DEMO-TF first winding pack proposal \cite{corato2022}), resulting in a total of 2280 nodes. The distance between each nodes is in this example of about 5 m, resulting therefore in a relatively coarse approximation. By scaling up to 15 N (x turns x layers) the latter would reduce to 3.3 m, while with 45 nodes to 1 m.  In the zoomed inset, examples of turn-turn bridges \rt and layer-layer bridges \rl are also shown, in purple and yellow, respectively.\\
It is possible to calculate in-code (few seconds) the mutual inductance matrix M as a result of the given geometry only (number of N, layers and turns) and joint and/or resistive bridge amount/position. This is done relatively fast (2-15 seconds depending on the total N/bridges number) without the longer workflow of codes typical of other commercial software. This allows - after optimization-  to set e.g. parametric simulations of different PI scenarios without manually transferring different input/output data.
\noindent Electro-magnetic computation is based on Kirchhoff’ current and voltage laws. The current distribution within the coil winding is obtained for any given temperature distribution by solving Kirchhoffs at each branch formed by transverse resistance (e.g. integral form of Maxwell equations). \\
At the same time, the thermal evolution is computed with the heat equation evaluated at each node, as displayed in Eq. \ref{heat}:
\begin{equation}
\label{heat}
C \frac {\delta T}{\delta t}= q_{joule}+\frac {\delta }{\delta x}\left(k \frac {\delta T}{\delta x} \right) +q_{transv}+ q_{ext}~(W/m^3)
\end{equation}\\
, where $T$ is temperature, $t$ is time, $C$ the heat capacity, $k$ the thermal conductivity, $q_{joule}$ the Joule heating from any resistive element, $q_{transv}$ the heat shared transversally by conduction and $q_{ext}$ whatever possible heat source from outside the coil winding (heaters, AC losses, etc). The current transfer between SC and stabilizer follows Eq. \ref{coupling}, with the assumption of equal electric EF field along the length (between nodes), while the transverse EF between domains is neglected (conductor cross-section):
\begin{equation}
\label{coupling}
\left( \frac {I_{sc}}{I_c}\right)^n+\frac {\eta_{st}I_{sc}}{A_{st}E_0}\left(\frac {I_{sc}}{I_c}-\frac {I}{I_c}\right)=0
\end{equation}
Where $\eta_{st}$ is the stabilizer resistivity, $A_{st}$ its cross section, $I_{sc}$ the current flowing in the strands/tapes bundle and $I_{c}$ their critical current. Taking again the example of N=10 (x layers x turns) based geometry (5 m inter-nodal distance), one can see that holding the hypothesis of equal electric field between nodes can be not realistic e.g. during a quench event in the nearby of the hot-spot. The amount of nodes has therefore be tuned according to the type of simulation one needs to run, by eventually optimizing the nodes number e.g. close to selected quench ignition point in the coil. Finally, the material properties database used in this code is the same as used in CryoSoft \cite{bottura2000}.  
\vspace{-5mm}
\subsection{Benchmarks}\label{benchmarks}
\begin{figure}[!t]
\centering
\includegraphics[width=0.8\textwidth]{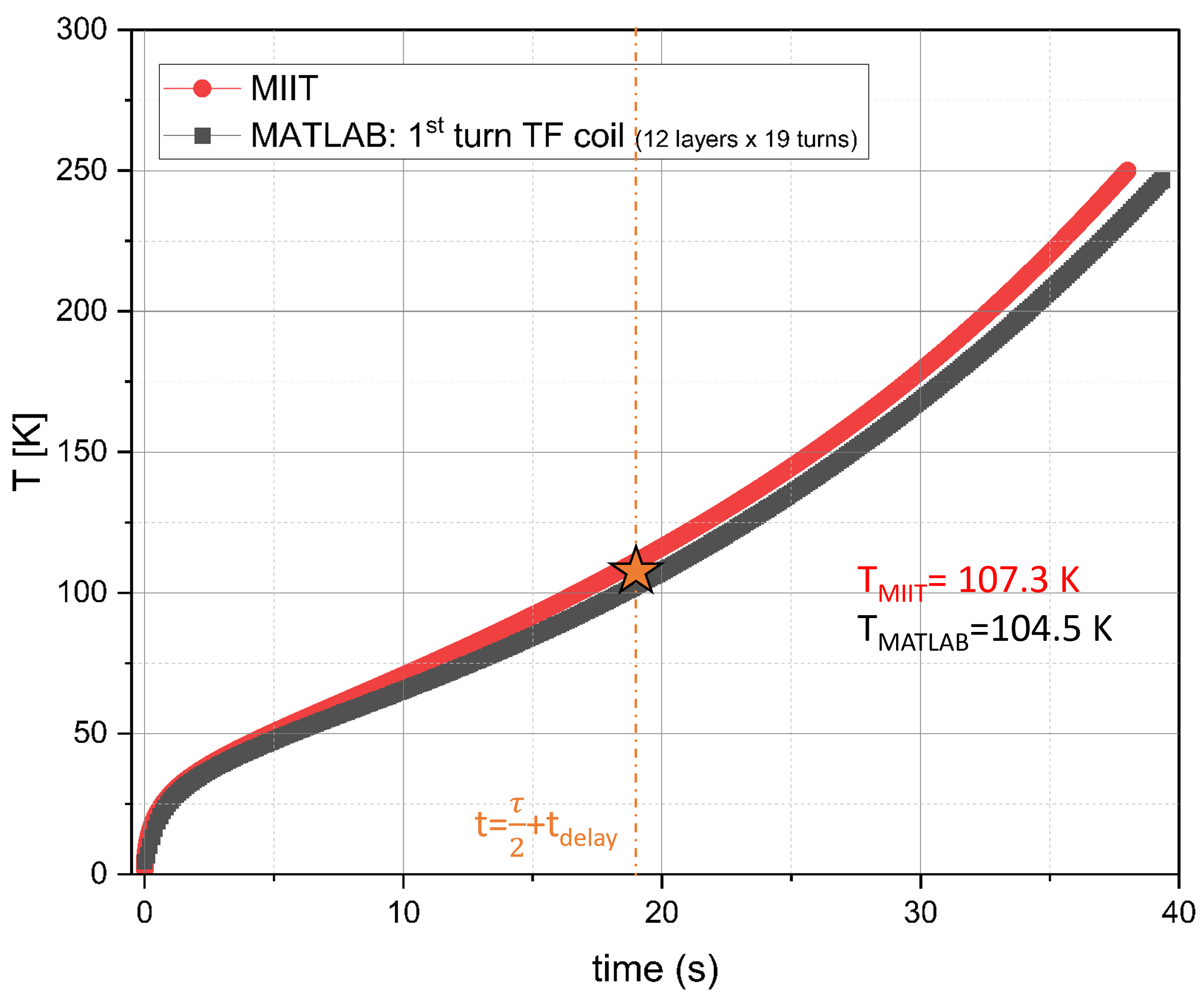}
\caption{Comparison between the temperature evolution (after a quench) for MIIT analytical expression and MATLAB simulation of a full TF DEMO geometry, fully insulated (kapton, fiber glass).}
\vspace{-4mm}
\label{benchmark2}
\end{figure}

\noindent The code quality was verified by undertaking some benchmark verifications. The first positive feedback comes from the DEMO TF PI-coil simulation cases compared to LTSpice\textregistered~ones, being in fact also possible to select the solely electrical computation in the MATLAB code (decoupling from thermal). In a DEMO TF configuration as from Tab.\ref{Tab2}, a current ramp that reaches I=66 kA in 30.000s is simulated for a coil where both layers and turns are shortened as following the output data of Sec. \ref{analytical}: the MATLAB code follows very well I and V (both inductive and resistive) evolutions at each loop/node, respecting the charging time delays for each of the turns and the voltage reached in any branch as in Spice.\\
Another relevant feedback is also obtained from the reproduction of the MIIT (''Mega-Current-Current-Time'') equation, which represents the full adiabatic case \cite{salmi2018}. This is generally used for having an idea of \Tmax~the (standard) insulated coil will have at the I-extraction characteristic time \cite{bajas2015}. In MATLAB, this was simulated by placing a heater in the first turn of a full simple geometry configuration (Tab. \ref{Tab2}), igniting a quench that generates a temperature evolution that follows the one displayed in black in Fig.\ref{benchmark2}. The field is set to B=12 T, with RRR=75.
The MATLAB code follows quite well the expected MIIT trend: the small difference between the curves stems from the fact that thermal conductivity terms $k$ can't be removed from the MATLAB model, while indeed the MIIT approach is strictly adiabatic. If one takes e.g. the characteristic extraction time of $t=\tau/2 + t_{delay} \simeq$19 s (typical $\tau$ for DEMO TF is about 35 s), this results in a difference of about 3 K. \\
Further benchmarks with other multi-physics software as ANSYS or COMSOL are to be exploited and would definitely endorse the validity of the model. Nevertheless, a true validation of the model can only come by comparing with experimental facts, the latter being a solid outlook for this work.
 \begin{table}[htbp!]
\begin{center}
\caption{Main parameters used for PI-DEMO TF numerical model benchmarks (Subsec.\ref{benchmarks}) and parametric study (Sec.\ref{results}).}
\label{Tab2}
\begin{tabular}{| c | c |}
\hline
Coil geometry  & N= 10 x 12 (layers) x 19 (turns) \\ 
\hline
Coil Inductance  & 3.55 H \\ 
\hline
node-node distance & 5 m  \\
\hline
B & 12 T  \\
\hline
RRR & 75  \\ 
\hline 
R\textsubscript{joint}   & 1 n$\Omega$  \\
  \hline 
Conductor cross section  & 2760$\cdot$10\textsuperscript{-3} m\textsuperscript{2};
\\ 
  & SS=72.20\%;SC=3.61\%; Cu=23.80\%;\\
\hline
\end{tabular}
\end{center}
\end{table}
\vspace{-7mm}
\section{First Results}\label{results}
\begin{figure}[htbp!]
\centering
\includegraphics[width=0.8\textwidth]{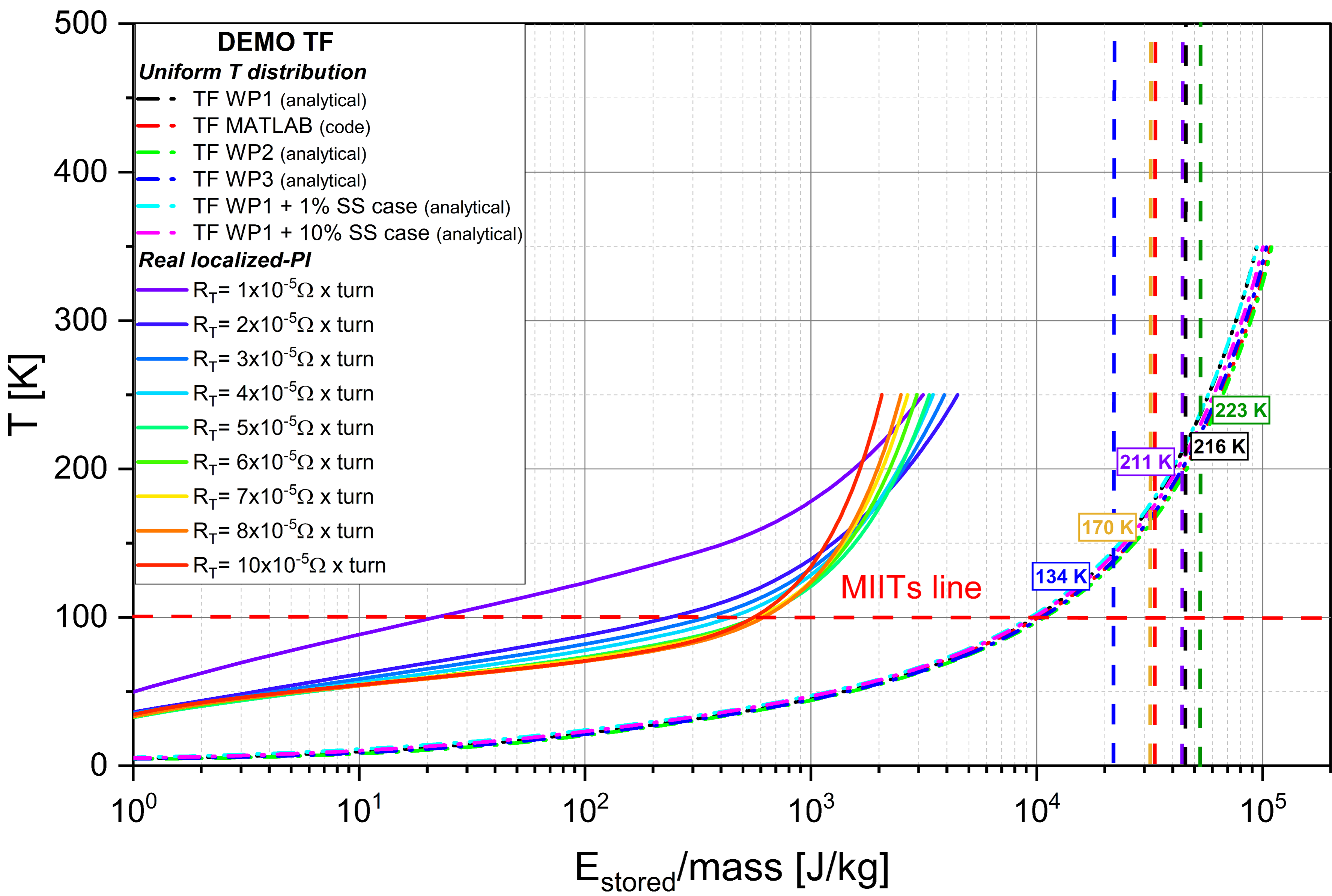}
\caption{Energy density (J/kg) vs temperature (K) trends for complete internal damping: dash-dot for different DEMO TF winding pack WP ideal cases (\cite{corato2022}) and solid lines for the simulated turn-turn localized PI case. Vertical lines indicate the exact energy densities for each of the DEMO WP entries.}
\vspace{-4mm}
\label{Results}
\end{figure}

\begin{figure}[htbp!]
\vspace{-4mm}
\centering
\includegraphics[width=0.8\textwidth]{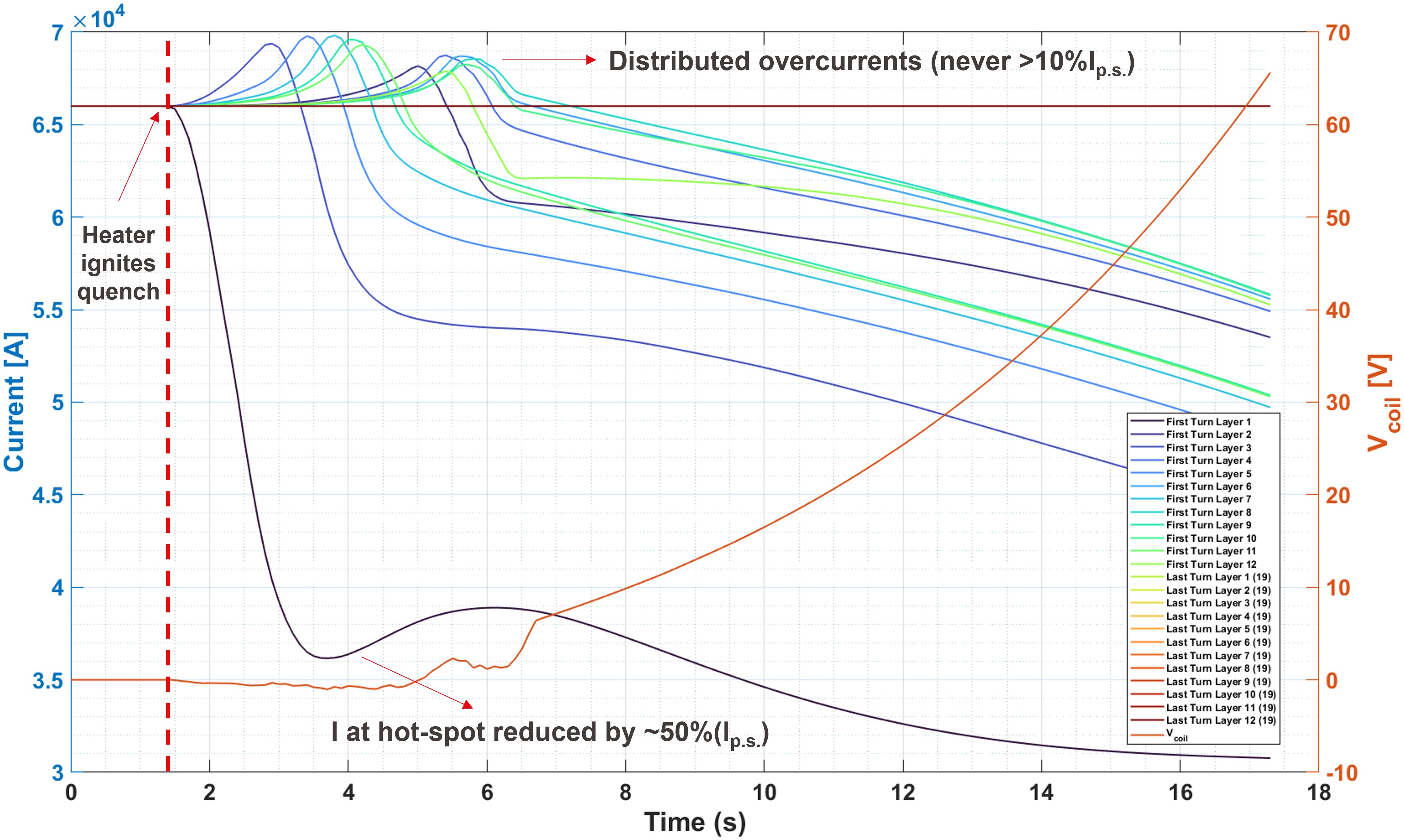}
\caption{Time vs Current/Voltage for a selected DEMO TF PI-coil case: \rt= 3 x 10\textsuperscript{-5}$\Omega$ x turn. The current from power supply I\textsubscript{p.s.} is not damped and keep staying constant throughout the quench event.}
\label{I-V}
\end{figure}

Parametric quench simulations were provided by starting with only turn-turn shortened DEMO TF coil, thus keeping the traditional insulation between layers. One has to think about these \rt~as localized rivet/screw alike metal connections placed at selected nodes, piercing the insulation layer (across its thickness, see purple line in Fig. \ref{Geometry}) that would otherwise be continuos also between adjacent turns of the magnet. \\
To have a first idea of \Tmax for DEMO TF coils, one can approach the ideal cases discussed in Sec. \ref{analytical} calculating it from their own energy densities - again via Eq. \ref{EvsM} - for different winding pack proposals (WPs) as in \cite{corato2022}: depending on their weight/material partition, \Tmax will range between $\simeq$130 K and $\simeq$ 225 K. These temperatures are already quite higher if compared with the typical T reached by the coil if one extracts the current in external damp resistors (see $\simeq$105 K in Fig.\ref{benchmark2}).\\
These results are summarized in Fig. \ref{Results}, where the energy density (J/kg) vs temperature (K) is shown for the mentioned cases and other two with more SS assumed to contributing heat exchange  (coloured dash-dotted lines all collapsing almost on the same one). In this plot, the vertical dashed lines represent the energy densities for each of the analysed cases (the exact points at which all the energy has been released) while the horizontal red-dashed is the MIIT line.\\
Simulations results (configuration as in Tab. \ref{Tab2}), with only \rt~active (total=1500 bridges), show instead a different scenario, being it also displayed in Fig. \ref{Results} (results stop at 250 K as simulation exit-condition). Initial values for \rt~were taken from Sec. \ref{analytical} (1$\mu\Omega$ x turn) and then incremented by distributing the same \rt~at different nodal planes (each distant 5 m). With this N/\rt~configuration, the best results (1-3$\cdot10^{-5}\Omega$ x turn) show extrapolated \Tmax~no lower than 400 K, being this not tolerable as for damage risks and long re-cooling process. Moreover, distributing more \rt~at farer nodes seems to even to worsen \Tmax. By not varying \rt, \Tmax~could be possibly mitigated by making the contact less localized, spreading it along the conductor length by increasing the number of nodes and bridges number. At the same time, new \rt~range should be tried, increasing the resistance from the minimum level established in Sec. \ref{analytical}.\\
Contrarily, the desired effect on current and voltage seems to be consistently obtained through each of the simulations, as it is displayed in the example of Fig. \ref{I-V}. Even if I=66 kA is not extracted in any of the simulations (keep staying there after the quench is always ignited by a heater at 1.5 s), the current at the hot-spot (first turn, black solid line) decreases by about 50\%  in about 2 s, while all the other neighbouring turns take part of it with a small over-current. This behaviour of current also benchmarks the results found in \cite{keilin2001}, while at the same time a reduction of \Vc is observed of about 3 orders of magnitude from the insulated case (e.g. 60 V max in Fig.\ref{I-V}). The latter gets decreased by increasing \rt~spot locations.
As for lowering \Tmax, another plausible solution is to introduce layer-layer bridges \rl, thus shortening the available path the heat has to propagate across the winding pack cross section. By keeping the same geometry there as in Tab. \ref{Tab2} but switching off this time \rt, \rl=2 m$\Omega$ were simulated with T=250 K being reached later, about 20\% beyond in time than the best achieved result with only \rt~active.  

\vspace{-3mm}
\section {Conclusions} 
\noindent The problem of internal energy damp in large-\Est SC magnets was discussed, particularly focusing on the EUROFusion DEMO TF case approached both analytically and with LTSpice simulations. As the results are very limiting, an original numerical model was developed producing coupled electro-thermal simulations. It was used to simulate the quench evolution of a NI/PI coil, already undergoing some benchmarks. A parametric study of DEMO TF PI case - where only localized metal-like contact resistances \rt~are accounted -  shows very high \Tmax~for complete internal damping ($\simeq$450 K), while \Vc stays very low (10-80 V) and currents at the hot-spot is reduced by almost 50\% (at the expenses of a small over-current). High \Tmax~can be mitigated by spreading the \rt~contact area, as well as introducing \rl~for a shorter thermal path, being this the outlook for future design work.

\end{document}